\documentclass[prl,twocolumn,preprintnumbers,amsmath,amssymb]{revtex4}
\usepackage{amsmath,amssymb,epsfig}
\usepackage{times}
\usepackage{latexsym}
\usepackage{multirow}
\usepackage{url}
\usepackage{color}
\usepackage{verbatim}

\usepackage{graphicx}
\usepackage{epsfig}
\usepackage{amssymb}
\usepackage{color}

\newcommand{\bal}{\begin{aligned}} \newcommand{\eal}{\end{aligned}}

\def\IR{\mathbb{R}}

\def\IZ{\mathbb{Z}}

\def\e{\mathrm{e}}

\def\clap#1{\hbox to 0pt{\hss#1\hss}}

\newcommand{\sect}[1]{\noindent\textit{#1}.---\!}
\renewcommand\tilde\widetilde

\renewcommand{\e}{\epsilon}

\newcommand{\bL}{\bar{{\mathcal{L}}}}

\newcommand{\be}[1]{ \begin{equation}\label{#1} }
\newcommand{\ee}{\end{equation}}
\newcommand{\ben}[1]{\begin{eqnarray}\label{#1} }
\newcommand{\een}{\end{eqnarray}}

\newcommand{\p}{\partial}

\newcommand{\refb}[1]{(\ref{#1})}

\renewcommand{\>}{\rangle}

\begin{document}

\thispagestyle{empty}

\title{Holography of 3d Flat Cosmological Horizons}
\author{Arjun Bagchi$^{a}$, St\'ephane Detournay$^{b}$, Reza Fareghbal$^{c}$, Joan Sim\'on$^{a}$}
\affiliation{
(a) School of Mathematics \& Maxwell Institute, University of Edinburgh, Edinburgh EH9 3JZ, UK.\\
(b)  Center for the Fundamental Laws of Nature, Harvard University, Cambridge, MA 02138, USA. \\
(c) School of Particles and Accelerators, Institute for Research in Fundamental Sciences (IPM), P.O. Box 19395-5531,Tehran, Iran.
}

\begin{abstract}
We provide a first derivation of the Bekenstein-Hawking entropy of 3d flat cosmological horizons in terms of the counting of states in a dual field theory. These horizons appear in the shifted-boost orbifold of $\IR^{1,2}$, the flat limit of non-extremal rotating BTZ black holes. These 3d geometries carry non-zero charges under the asymptotic symmetry algebra of $\IR^{1,2}$, the 3d Bondi-Metzner-Sachs (BMS${}_3$) algebra. The dual theory has the symmetries of the 2d Galilean Conformal Algebra, a contraction of two copies of the Virasoro algebra, which is isomorphic to BMS${}_3$. We study flat holography as a limit of AdS${}_3$/CFT${}_2$ to semi-classically compute the density of states in the dual, exactly reproducing the bulk entropy in the limit of large charges. Our flat horizons, remnants of the BTZ inner horizons also satisfy a first law of thermodynamics. We comment on how the dual theory reproduces the bulk first law and how cosmological bulk excitations are matched with boundary quantum numbers.
\end{abstract}

\maketitle

\renewcommand{\thepage}{\arabic{page}}
\setcounter{page}{1}
\def\thefootnote{\arabic{footnote}}
\setcounter{footnote}{0}

\sect{Introduction} 
The holographic principle \cite{'tHooft:1993gx, Susskind:1995vu} and its explicit realization in terms of the Anti de Sitter/Conformal Field Theory (AdS/CFT) correspondence \cite{Maldacena:1997re} are among the most successful conceptual frameworks for quantum gravity. The understanding of holography in general spacetimes and of the microscopic nature of entropy for general horizons is crucial for a further development of this field. In this letter, we report some progress on holography for three dimensional (3d) asymptotically flat spacetimes and on the microscopic counting for the entropy of cosmological horizons in this theory. 


Flat spacetimes can be obtained as large radius limits of AdS. Formulating flat space holography as a limit of AdS/CFT has been used to extract features of the flat space S-matrix from AdS/CFT correlators, see e.g.~\cite{Polchinski:1999ry, Susskind:1998vk, Giddings:1999jq, Gary:2009ae}. Our 3d analysis follows this philosophy and is primarily based on the symmetries of the theory. This was recently pursued in \cite{Barnich:2010eb,Bagchi:2010zz,Bagchi:2012cy,Barnich:2012aw} where these flat limits are realized as a contraction on the symmetry structure of AdS/CFT. On the bulk side \cite{Barnich:2006av,Barnich:2010eb,Barnich:2012aw}, the AdS symmetries are contracted to what is called the Bondi-Metzner-Sachs (BMS) algebra 
\cite{Bondi:1962,Sachs:1962,Sachs:1962wk,Ashtekar:1996cd}. On the field theory side, the CFT symmetries are contracted to the so called Galilean Conformal Algebras (GCA), previously considered in the context of non-relativistic limits of CFTs \cite{Bagchi:2009my}. This connection was christened the BMS/GCA correspondence \cite{Bagchi:2010zz}, which holds even beyond our 3d context.

In this letter we investigate the correspondence in 3 bulk dimensions and explore the extent to which the dual 2d Galilean Conformal field theory (GCFT), invariant under GCA, can reproduce non-trivial features about asymptotic $\IR^{1,2}$ spacetimes. We show the family of {\it shifted- boost orbifolds} of $\IR^{1,2}$ \cite{Cornalba:2002fi}, that correspond to the flat limit of the non-extremal Banados-Teitleboim-Zanelli (BTZ) black holes \cite{Banados:1992wn,Banados:1992gq}, carry non-trivial BMS${}_3$ charges.
Their causal structure contains a cosmological horizon, the remnant of the BTZ inner horizon, whose area, surface gravity and rotation allow us to define a bulk entropy, temperature and angular velocity, respectively. We show these satisfy a first law of thermodynamics, analogous to the ones recently discussed in the literature \cite{Cvetic:1997uw,Park:2006pb, Ansorg:2008bv,Ansorg:2009yi,Cvetic:2010mn,Castro:2012av,Detournay:2012ug,Chen:2012mh}.   

We then derive an analogue of Cardy's formula for 2d GCFTs by computing their density of states semi-classically. Our main result is that our dual field theory calculation reproduces the bulk entropy. This constitutes the first ``microscopic" derivation of the entropy of a cosmological horizon, in the spirit of \cite{Strominger:1997eq}. We further point out 2d GCFTs have negative thermodynamic temperatures and specific heats. We conclude by matching the quantum numbers and frequencies measured by a bulk cosmological observer with the ones in the 2d GCFT.

\sect{Asymptotic Symmetries in 3d Minkowski Spacetime} The physical states in the Hilbert space of a quantum theory of gravity form representations of the symmetry structure of the theory at its asymptotic boundary, the asymptotic symmetry group and its corresponding algebra, the asymptotic symmetry algebra (ASA). In AdS$_3$, the seminal work of \cite{Brown:1986nw} showed that the ASA is formed by two commuting copies of the Virasoro algebra $( \mathcal{L}_n, \bL_n)$
\be{vir}
[ \mathcal{L}_m, \mathcal{L}_n] = (m-n) \mathcal{L}_{m+n} + \frac{c}{12} m(m^2 -1) \delta_{m+n,0}\,,
\ee
with analogous brackets for $\bL_n$, and central charge $\bar c=c = \frac{3 \ell}{2G}$, where $\ell$ is the radius of AdS$_3$ and $G$ the Newton constant. These Virasoro algebras are also physically realized as the local symmetry algebras of 2d CFTs. The two are identified, through the AdS/CFT correspondence \cite{Maldacena:1997re}, providing a match between the gravity and the dual CFT symmetries. 

In 3d Minkowski spacetimes, the ASA is given by the BMS${}_3$ algebra \cite{Barnich:2006av}
\ben{bms3}
[L_m, L_n] = (m-n) L_{m+n} + c_{\textrm{\tiny LL}} m(m^2 -1) \delta_{m+n,0},  &&\crcr 
[L_m, M_n] = (m-n) M_{m+n} +  c_{\textrm{\tiny LM}} m(m^2 -1) \delta_{m+n,0} &&
\een
with $[M_n,\,M_m]=0$ and $c_{\textrm{\tiny LL}}=0, c_{\textrm{\tiny LM}}= 1/4$. These symmetries are defined at null infinity, with $L_m$ being the diffeomorphisms of its spatial circle and $M_n$ the angular dependent supertranslations and translations. This algebra is isomorphic to the 2d GCA, the symmetry algebra of GCFTs, which are best viewed as a limit of standard 2d CFTs. This forms the basis of the BMS/GCA correspondence in 3 bulk dimensions. The precise map can be described in terms of a In{\"o}n{\"u}-Wigner contraction of the two copies of the Virasoro algebra \cite{Barnich:2006av}
\be{vir2gca}
L_n={\mathcal{L}}_n-\bar{\mathcal{L}}_{-n}, M_n={\e}\left({\mathcal{L}}_n+\bar{\mathcal{L}}_{-n}\right), \, \e= {G}/{\ell} \to 0
\ee
where we take $\ell \to \infty$ keeping $G$ fixed. This limit implemented on the algebra \refb{vir} yields \refb{bms3},
with central charges $c_{\textrm{\tiny LL}} = \frac{1}{12}(c-\bar c)=0$ and $c_{\textrm{\tiny LM}} = \frac{G}{12\ell}(c+\bar c)=1/4$.

A natural spacetime interpretation in terms of the contraction of the Killing vectors of global AdS to the Killing vectors of flat-space and the natural extension to the infinite set of asymptotic Killing vectors was recently worked out in \cite{Bagchi:2012cy}, where it was also shown that the $\ell \to \infty$ limit induces a spacetime contraction $(t, x) \to (\e t, x)$ on the CFT living on the boundary cylinder. This limit is best interpreted as an ultra-relativistic limit {\footnote {One can look at an inequivalent contraction of the two Virasoros $L_n= \mathcal{L}_n + \bL_n, \, M_n={\e}(\mathcal{L}_n - \bL_n)$ \cite{Bagchi:2009pe} which can be understood as a non-relativistic $(x, t) \to (\e x, t)$ contraction in terms of spacetime. This yields the same algebra \refb{bms3}. We will continue calling \refb{bms3} GCA$_2$ in the dual theory to be consistent with previous literature. \label{fn1}}}.

\sect{Spacetime analysis}
In the absence of sources, the most general solution to 3d pure gravity with vanishing cosmological constant is locally flat \cite{Deser:1983tn} and it can be written as \cite{Barnich:2010eb}
\begin{equation}\label{bms3metric}
 ds^2 = \Theta(\psi)du^2  -2drdu + 2[\Xi(\psi) + \frac{u}{2}\Theta^\prime] d\psi du + r^2 d\psi^2 .
\end{equation}
The asymptotic structure at null infinity is preserved by a set of diffeomorphisms depending on two arbitrary functions of $\psi$ 
whose modes \cite{Barnich:2006av,Bagchi:2012yk} 
\begin{subequations}
\label{eq:asydiff}
\begin{align}
 \ell_n &= i e^{i n \psi} \big(i n u \p_u - i n r\p_r +(1+ n^2 \frac{u}{r}) \p_\psi)\\ 
 m_n &= i e^{i n \psi} \p_u\,, \quad \quad n\in\IZ
\end{align}
\end{subequations}
satisfy the centerless BMS${}_3$ algebra. Note that $\ell_{\pm 1,0}$ and $m_{\pm 1,0}$ coincide asymptotically with the exact Killing vectors of Minkowski space forming the global $\mathfrak{iso}(2,1)$ subalgebra of BMS$_3$.
The coefficients of the Fourier mode decomposition of the corresponding asymptotically conserved charges $L_n$ and $M_n$ determine the arbitrary functions $\Theta(\psi)\,,\Xi(\psi)$:
\begin{equation}
 \Theta = -1 + 8G\sum_n M_n\,e^{-in\psi}\,, \quad \Xi = 4G\sum_n L_n\,e^{-in\psi}\,.
\end{equation}
In this letter, we study the most general zero mode solution labelled by $\Theta=8GM$ and $\Xi=4GJ$ with $M,\,J\geq 0$. We show these correspond to the {\it shifted-boost orbifold} $(J\neq 0)$ \cite{Cornalba:2002fi} and the boost orbifold $(J=0)$ \cite{Khoury:2001bz} of $\IR^{1,2}$.
This claim can be derived by taking the flat limit, as in \cite{Cornalba:2002fi}, of the non-extremal BTZ black holes $(M\ell \neq J)$, 
\begin{eqnarray}
\label{eq:btz}
  ds_{\textrm{\tiny BTZ}}^2&=&(8 G M-\frac{r^2}{\ell^2}) dt^2 + \frac{dr^2}{-8 G M + \frac{r^2}{  \ell^2} + \frac{16 G^2 J^2}{r^2}} \cr
  & &- 8 G J dt d\phi + r^2 d\phi^2,\quad \quad \phi\sim \phi + 2\pi\,.
\end{eqnarray}
whose outer and inner horizons $r_\pm$ are given by 
\begin{eqnarray}
  M = \frac{r_+^2 + r_-^2}{8 G \ell^2}, \quad J =\frac{r_+ r_-}{4 G \ell}.
\end{eqnarray}
We refer to their flat $\ell\to\infty$ limit as the ``flat-BTZ" (FBTZ)
\ben{eq:shiftorb}
&&ds_{\textrm{\tiny FBTZ}}^2=\hat r_+^2 d t^2-\frac{r^2\,dr^2}{\hat r_+^2 (r^2-r_0^2)}+r^2 d\phi^2-2 \hat r_+ r_0
 dt d \phi, \;\;\quad
 \een
where $r_+ \to \ell \sqrt{8GM}  = \ell \hat{r}_+$ and $r_-\to \sqrt{\frac{2G}{M}} |J| = r_0$.
These correspond to \eqref{bms3metric} under the coordinate transformation
\begin{equation*}
  d\psi =d\phi + \frac{r_0\,dr}{\hat r_+\,(r^2-r_0^2)},\,\,\,\,
  du = dt + \frac{r^2\,dr}{\hat r_+^2\,(r^2-r_0^2)} .
\end{equation*}

The null hypersurface $r=r_0$ is a Killing horizon with normal $\chi=\partial_u +(\hat r_+/r_0)\partial_\phi$, surface gravity $\kappa = \hat r_+^2/r_0$ and angular velocity $\Omega = \hat r_+/r_0$. Thus, we can associate a Hawking temperature and entropy to it
\begin{equation*}
  T_{\textrm{\tiny FBTZ}} = \frac{\kappa}{2\pi} = \frac{\hat r_+^2}{2\pi\,r_0},\,\,\, S_{\textrm{\tiny FBTZ}} = \frac{\pi\,|r_0|}{2G}\,.
\end{equation*}
In fact, these quantities satisfy the first law of thermodynamics
\begin{equation}\label{eq:flawbulk}
  -T_{\textrm{\tiny FBTZ}} \, dS_{\textrm{\tiny FBTZ}} = dM - \Omega\,dJ
\end{equation}
which is the remnant of the corresponding law satisfied by the BTZ inner horizon at finite $\ell$
\begin{equation}
\label{eq:bulk1law}
  -T_H^-dS^- = dM - \Omega^- dJ ,
\end{equation}
where $T_H^- = \frac{r_+^2-r_-^2}{2\pi r_-\ell^2}\,,\,\, \Omega^- = \frac{r_+}{\ell r_-}\,,\,\,S^- = \frac{\pi |r_-|}{2G_3}$
are its temperature, angular velocity and entropy, respectively. 

That \eqref{eq:shiftorb} is a quotient manifests through the map to cartesian coordinates (valid for $r>r_0$, a similar change of coordinates exists for $r<r_0$):
\begin{eqnarray}
  X^2 &=& \frac{r^2 - r_0^2}{\hat r_+^2}\,\sinh^2(\hat r_+\,\phi) \quad T^2 = \frac{r^2 - r_0^2}{\hat r_+^2}\,\cosh^2(\hat r_+\,\phi) , \cr
  Y &=& r_0\,\phi - \hat r_+\,t 
\label{eq:c1}
\end{eqnarray}
so that $\xi = \partial_\phi = r_0 \partial_Y + \hat r_+\,(X\partial_T + T\partial_X)$ acts like $X^\pm\sim e^{\pm2\pi\hat r_+}\,X^\pm,\,\,Y\sim Y + 2\pi r_0$. Equivalently, the action of $\xi$ identifies points of $\IR^{1,2}$ under a combined boost in the $(T,X)$ $\IR^{1,1}$ plane of rapidity $\hat{r}_+$ and a translation of length $r_0$ in the transverse Y-direction. Thus, \eqref{eq:shiftorb} is the shifted-boost orbifold of $\IR^{1,2}$ \cite{Cornalba:2002fi}.

This orbifold interpretation provides a global description for the spacetime \refb{eq:shiftorb} \cite{Cornalba:2003kd}. Whenever $X^+X^- > 0$, 
\begin{equation}
  X^\pm = \frac{\tau}{\sqrt{2}}\,e^{\pm E(z+y)},\,\,\, Y = z,\,\,\,E=\frac{\hat r_+}{r_0}
\label{eq:c2}
\end{equation}
the geometry describes a contracting universe $(\tau<0)$ towards a cosmological horizon located at $\tau=0$, i.e. $r=r_0$, and an expanding one $(\tau>0)$ from it
\begin{equation}
  ds^2 = -d\tau^2 + \frac{(E\tau)^2}{H^2} dy^2 + H^2\left(dz + A\right)^2
\end{equation}
with $H^2 = 1 + (E\tau)^2$ and $A = (1-H^{-2})dy$. In the region $-1/E^2 < 2X^+X^- < 0$, 
\begin{equation}
  X^\pm = \pm \frac{x}{\sqrt{2}}\,e^{\pm E(z+y)},\,\,\, Y = z
\end{equation}
the geometry is static
\begin{equation}
  ds^2 = dx^2 - \frac{(Ex)^2}{H^2} dy^2 + H^2\left(dz + A\right)^2
\end{equation}
with $H^2 = 1 - (Ex)^2$ and $A = (1-H^{-2})dy$ and describes a Rindler space in the region $(Ex)^2\ll 1$ with a Rindler horizon at $x=0$, i.e. $r=r_0$. Finally, whenever $2X^+X^- < -1/E^2$, the geometry has closed timelike curves, as non-extremal BTZ black holes do. As in that case, we will excise this region from spacetime, introducing a singularity at its boundary $2X^+X^-=-1/E^2$.\\

\vspace{-0.4cm}

\sect{Dual field theory analysis} 
Quantum gravity states in 3d flat space should transform under representations of the infinite 2d GCA. 
These are labelled by eigenvalues of $L_0, M_0$ \cite{Bagchi:2009ca,Bagchi:2009pe}:
\ben{hlhm}
&&L_0 | h_L, h_M \> = h_L | h_L, h_M \>, \, M_0 | h_L, h_M \> = h_M | h_L, h_M \> \nonumber \\
&&\mbox{where} \,\  h_L = \lim_{\e \to 0} (h - \bar h), \quad   h_M = \lim_{\e \to 0} \epsilon (h + \bar h) 
\een
There exists the usual notion of primary states $| h_L, h_M \>_p$ in the 2d GCA annihilated by $L_{n}, M_{n}$ for $n>0$. Representations are built by acting with the raising operators $L_{-n}, M_{-n}$ on them. Thus, they satisfy $h_L \geq 0$. Using the dictionary between $h,\,\bar h$ and AdS${}_3$ mass and angular momentum, we can relate $\{h_L,\,h_M\}$ to the BMS${}_3$ charges
\ben{FBTZ-charges}
&& h = \frac{1}{2}( \ell M + J) + \frac{c}{24}, \quad \bar{h} = \frac{1}{2}( \ell M - J) + \frac{\bar{c}}{24} \cr
&& \Rightarrow h_L = J, \quad h_M = GM + \frac{c_{\textrm{\tiny LM}}}{2} = GM + \frac{1}{8}\,.
\een
This suggests the bound $h_M\geq 0$, saturated by $\IR^{1,2}$ \cite{Barnich:2012aw} given that $GM=-1/8$ for global AdS${}_3$. 
This is confirmed by a 2d GCFT unitarity bound derived from demanding the norm of a state of weight $(h, \bar h)$ at a level $n$ be non-negative in the original 2d CFT. This gives  $2 n h+ \frac{c}{12} n(n^2 -1) \geq 0$, and similarly for $\bar h$. Using the definitions \refb{hlhm}, one can derive the analogous statement for 2d GCFT
\be{uni-gca}
n h_M + c_{\textrm{\tiny LM}} n(n^2-1) \geq 0 \Rightarrow h_M \geq 0, c_{\textrm{\tiny LM}} \geq 0\,.
\ee
A more thorough analysis is required to better understand aspects of unitarity in 2d GCFTs.

Given the success of Cardy's formula in accounting for the entropy of BTZ black holes \cite{Strominger:1997eq}, it is natural to wonder whether a counting of states in 2d GCFTs can reproduce the gravitational entropy $S_{\textrm{\tiny FBTZ}}$. To analyze this, define the partition function of 2d GCFT as
\be{pf}
Z_{\textrm{\tiny GCFT}} (\eta, \rho) = \sum d(h_{\textrm{\tiny L}}, h_{\textrm{\tiny M}}) e^{2 \pi i (\eta h_{\textrm{\tiny L}}+\rho h_{\textrm{\tiny M}})}
\ee
where $d(h_L, h_M)$ is its density of states with charges $\{h_L,\,h_M\}$. To derive an analogue of Cardy's formula for the GCFT, it is crucial to use an analogue of modular invariance in the original 2d CFT for the 2d GCFT partition function. In particular, we need to derive the S-transformation rules for 2d GCFTs. We shall first state this result and then motivate it as emerging as a limit of the original 2d CFT. 
The S-transformation in 2d GCFTs reads 
\be{eq:unimod}
S: (\eta, \rho) \to (-{1}/{\eta}, {\rho}/{\eta^2}) 
\ee
To understand this, let us start with the 2d CFT partition function and rewrite it using \refb{hlhm} but at finite $\e$ 
\be{pf-cft}
Z_{\textrm{\tiny CFT}} =  \sum d_{{\textrm{\tiny CFT}}} (h, {\bar{h}}) e^{2 \pi i (\tau h+{\bar{\tau}} {\bar{h}})}= \sum d(h_{\textrm{\tiny L}}, h_{\textrm{\tiny M}}) e^{2 \pi i (\eta h_{\textrm{\tiny L}}+\frac{\rho}{\e} h_{\textrm{\tiny M}})} \nonumber
\ee
where $\eta,  \rho = \frac{1}{2}(\tau \pm \bar \tau)$. At finite $\e$, the S-transformation of the 2d CFT reads
\be{finiteS}
(\tau, \bar\tau) \to \left(- \frac{1}{\tau}, - \frac{1}{\bar \tau}\right) \Rightarrow (\eta, \rho) \to \left(\frac{\eta}{\rho^2 - \eta^2}, \frac{-\rho}{\rho^2 - \eta^2}\right).
\ee 
The hamiltonian scaling in $Z_{\textrm{\tiny CFT}}$ must be accompanied by a temperature rescaling $\rho \to {\e} \rho$. 2d GCFT transformations \refb{eq:unimod} emerge from \refb{finiteS} after this rescaling. Following \cite{Carlip:1998qw,Hotta:2010qi} closely,
we define the quantity
\be{}
Z_{\textrm{\tiny GCFT}}^0 (\eta, \rho) = e^{-\pi i \rho c_{\textrm{\tiny LM}}} Z_{\textrm{\tiny GCFT}} (\eta, \rho),
\ee
in analogy to the modular invariant 2d CFT partition function on the plane $Z^0_{\textrm{\tiny CFT}} = e^{-\frac{\pi i}{12}(c\tau - \bar c\bar \tau)}  Z_{\textrm{\tiny CFT}}$ satisfying $Z^0_{\textrm{\tiny CFT}} (\tau,\bar \tau) = Z^0_{\textrm{\tiny CFT}} (-1/\tau,-1/\bar \tau)$. In the flat limit, this translates into $Z_{\textrm{\tiny GCFT}}^0 (\eta, \rho) = Z_{\textrm{\tiny GCFT}}^0 (- \frac{1}{\eta},\frac{\rho}{\eta^2})$,
the S-transformation (\ref{eq:unimod}), leading to
\be{}
Z_{\textrm{\tiny GCFT}} (\eta, \rho) = e^{i \pi c_{LM} \rho (1- \frac{1}{\eta^2})} \,\ Z_{\textrm{\tiny GCFT}} \bigg(- \frac{1}{\eta},\frac{\rho}{\eta^2}\bigg) .
\ee
By doing an inverse Laplace transformation, the density of states equals
\ben{den}
d(h_L, h_M) = \int d \eta d \rho \,\ e^{2 \pi i {{f}}(\eta, \rho)} Z_{\textrm{\tiny GCFT}} (- 1/\eta, \rho/\eta^2)\\
\text{where} \,\ 
f(\eta, \rho) =  \left(\frac{c_{\textrm{\tiny LM}}}{2} - \frac{c_{\textrm{\tiny LM}}}{2\eta^2} -h_M\right)\,\rho - h_L \eta\,. \nonumber
\een
In the limit of large charges, we evaluate \refb{den} by a saddle-point approximation. There is a saddle at $\eta \approx i \sqrt{c_{\textrm{\tiny LM}}}/\sqrt{2h_M}$ whenever $Z_{\textrm{\tiny GCFT}} (- 1/\eta, \rho/\eta^2)$ is slowly varying, which occurs at positive $i\rho$, i.e. negative GCFT temperature, a point we stress below.
The 2d GCFT entropy is then 
\be{}
\label{eq:cardygca} 
S_{\textrm{\tiny GCFT}} = \log d(h_L, h_M) = 2\pi\,h_L\sqrt{\frac{c_{\textrm{\tiny LM}}}{2h_M}}. 
\ee

\vspace{-0.2cm}
This is the analogue of the Cardy formula for 2d GCFT \footnote{A similar Cardy-like GCA$_2$ formula was derived in the context of non-relativistic AdS/CFT in \cite{Hotta:2010qi}. Our derivation relies on a different contraction of the original CFT and is hence different. To arrive at the results of \cite{Hotta:2010qi}, one needs to use the scaling $\eta \to \e \eta$ in the S-transformation and follow the steps outlined above.}. Applying \refb{eq:cardygca} to the charges
\refb{FBTZ-charges} describing the shifted-boost orbifold (in the limit of large charges), one finds
\be{}
\boxed{
S_{\textrm{\tiny GCFT}} = \frac{\pi |J|}{\sqrt{2GM}} = S_{\textrm{\tiny FBTZ}}\,.
}
\ee
Thus, the 2d GCFT state counting exactly reproduces the entropy of the cosmological horizon.

Notice that 2d GCFT thermodynamic potentials equal
\begin{equation}
  \frac{\partial S}{\partial h_L} = \frac{\Omega^-}{T_H^-}\,, \quad
  \frac{1}{T_{\textrm{\tiny GCFT}}}\equiv \frac{\partial S}{\partial h_M} = -\frac{1}{G\,T_H^-}\,. 
\end{equation}
Thus, a universal feature of 2d GCFTs is the negativity of their thermodynamic temperature and specific heat
\begin{equation}
  C_M = \left.\frac{\partial h_M}{\partial T_{\textrm{\tiny GCFT}}}\right|_{h_L} = -\frac{\pi^2}{G}\,\frac{T_H^-}{(\Omega^-)^2}\,.
\end{equation}
Our Cardy formula \refb{eq:cardygca} is compatible with the bulk first law \refb{eq:bulk1law}, capturing its peculiar sign through the negative temperature, since $h_M\geq 0$ in 2d GCFTs. We view this as a very interesting feature of these theories given the negative specific heat that higher dimensional asymptotically flat black holes have, a feature that has always been difficult to reconcile with a dual field theory formulation.\\

\vspace{-0.4cm}

\sect{Cosmological Interpretation}
The shifted-boost orbifold \refb{eq:shiftorb} is only invariant under the translation $\partial_Y$ and the boost $X\partial_T + T\partial_X$. Its asymptotic structure at null infinity still satisfies the BMS${}_3$ boundary conditions, in the same way as BTZ black holes do preserve AdS${}_3$ asymptotics. Thus, an observer at null infinity can still assign non-trivial 2d GCA charges $M_0$ and $L_0$ to this geometry and use the 2d GCFT partition function to count the number of states carrying these charges.

A cosmological observer, who sees some contraction and expansion of the universe, will measure some temperature due to particle creation by the cosmological horizon. The latter is measured with respect to his cosmological clock. To discuss how this description is related to the one in 2d GCFT, consider a bulk scalar field excitation $\Phi$ with Fourier decomposition $\Phi = G_{pn}(\tau)\,e^{i\left(py + nz/r_0\right)}$. $p$ corresponds to the boost quantum number, whereas $n\in\IZ$ is momentum along the orbifold direction $z$. It was shown in \cite{Cornalba:2002nv} that excitations satisfying Dirichlet boundary conditions at the singularity have an asymptotic behaviour at $|\tau|\to \pm \infty$
\begin{equation}
  \Phi \to \frac{1}{\sqrt{\omega\,|\tau|}}\,e^{i\left(\pm \omega |\tau| + py\right)}\,e^{inz/r_0}\,,
\end{equation}
where for large charges (or massless fields) $\omega^2 = (p-n/r_0)^2$. Using \refb{eq:c1}-\refb{eq:c2}, one can relate this cosmological description to the BMS one as
\begin{equation}
  \Phi \to \sqrt{\frac{\hat r_+}{\omega r}}e^{i\omega r/\hat r_+}\,e^{i\left(\hat r_+\,t(p-n/r_0) + n\phi\right)}
\end{equation}
Thus, frequencies $\omega$ measured by the cosmological observer agree with those measured at infinity $(\hat r_+\,(p-n/r_0))$, up to the rescaling $\hat r_+$, which also ensures the matching of the Hawking temperature due to particle creation radiation with the surface gravity temperature $T_{\textrm{\tiny FBTZ}}$ \cite{Cornalba:2002nv}. \\

\vspace{-0.4cm}

\sect{Discussions}
In this letter we have advocated that an appropriate limit of the 2d CFT dual to $AdS_3$ potentially yields a field theory dual to 3d flat space.
Our formalism heavily relied on the symmetry structure of the theory. It would be interesting to push a more dynamical and operational meaning to flat holography in this context, for example, by understanding how to reproduce correlation functions in GCFTs \cite{Bagchi:2012cy} using a bulk analysis both at zero and finite temperatures. One could for instance consider string theory setups where the CFTs are identified and of a particular type to determine if a concrete realization of the proposal could be envisaged. 

Let us comment on an important property of 2d GCAs. The representations of $M_0$ are non-diagonal \cite{Bagchi:2009pe, Hotta:2010qi, Bagchi:2012yk}, a feature reminiscent of structures encountered in Logarithmic CFTs. Thus, unitarity in 2d GCFTs deserves a more careful analysis. This off-diagonal nature is also relevant for entropy considerations. In fact, the density of states in \refb{pf} is only a good approximation in the Cardy regime of large charges to the full partition function $Z_{\textrm{\tiny GCFT}} = \mbox{Tr} \, e^{2 \pi i \eta L_0} \, e^{2 \pi i \rho M_0}$. Outside this regime, the off-diagonal corrections to the Cardy-like formula will become important.

Our set-up is also interesting because allows us to address questions in cosmological backgrounds using a holographic description. There are open questions in this context. The shifted-boost orbifold has a classically stable Cauchy horizon \cite{Cornalba:2003ze} when its singularity is interpreted as an orientifold in string theory \cite{Cornalba:2002nv}. It would be desirable to have a quantum version of this statement when including coupling to matter. It is also natural to investigate whether these solutions can be interpreted as thermal states in the 2d GCFT dual, for instance, by computing their quasi-normal modes.

It is perhaps worth emphasizing that at finite $\ell$, $Z_{\textrm{\tiny CFT}}$ has exponentially suppressed contributions to the BTZ entropy of the form $\exp(S_- - S_+)$, with $S_+$ the entropy of the BTZ outer horizon. In the flat limit, one of them becomes the physically dominant one, reproducing the $\ell\to \infty$ entropy of the BTZ inner horizon. From this perspective, it would be interesting to explore whether the thermodynamics satisfied by 2d GCFTs is related to the {\it finite} $\ell$ inner horizon BTZ instabilities \cite{Steif:1993zv,Balasubramanian:2004zu}. 


\sect{Acknowledgements}
We are grateful to M. Alishahiha, D. Allahbakhshi, S. Carlip, A. Castro, R. Gopakumar, D. Grumiller, T. Hartman, D. Hofman, A. E. Mosaffa, G-S. Ng for useful discussions. SD thanks G. Barnich. SD was supported by the Fundamental Laws Initiative of the Center for the Fundamental Laws of Nature, Harvard University. This work was partially supported by the NSF under grant PHY05-51164, by EPSRC under grant EP/G007985/1 and by an Excellence Fellowship of Wallonie-Bruxelles International.

\vspace{-0.6cm}


\end{document}